# An Improved Robust Total Logistic Distance Metric algorithm for Generalized Gaussian Noise and Noisy Input

Haiquan Zhao, *Senior Member*, *IEEE*, Yi Peng, Zian Cao

*Abstract*—Although the known maximum total generalized correntropy (MTGC) and generalized maximum blake–zisserman total correntropy (GMBZTC) algorithms can maintain good performance under the errors-in-variables (EIV) model disrupted by generalized Gaussian noise, their requirement for manual adjustment of parameters is excessive, greatly increasing the practical difficulty of use. To solve this problem, the total arctangent based on logical distance metric (TACLDM) algorithm is proposed by utilizing the advantage of few parameters in logical distance metric (LDM) theory and the convergence behavior is improved by the arctangent function. Compared with other competing algorithms, the TACLDM algorithm not only has fewer parameters, but also has better robustness to generalized Gaussian noise and significantly reduces the steady-state error. Furthermore, the analysis of the algorithm in the generalized Gaussian noise environment is analyzed in detail in this paper. Finally, computer simulations demonstrate the outstanding performance of the TACLDM algorithm and the rigorous theoretical deduction in this paper.

## I. INTRODUCTION

DIGITAL signal processing has been developed for many years, and since the classical least mean squares (LMS) algorithm [1] was proposed, adaptive filtering algorithms are extensively available for the fields of system identification, echo cancellation, and frequency estimation for power systems [2][3][4]. In fact, the input signal often contains noise as well, due to personnel errors, sensor defects, and other problems. Yet, the LMS algorithm fails to consider the situation where the input signal is affected from noise as well. Therefore, classical adaptive filtering algorithms, such as the LMS algorithm, suffer from performance degradation [5] in EIV models [6].

The total least-squares (TLS) algorithm is one of the useful methods to cope with the situation where the input signal is affected from noise as well. The gradient descent TLS (GDTLS) algorithm proposed in [7] effectively addresses the above issue. On this basis, to further strengthen the robustness of the algorithm, some robust algorithms based on correlation entropy theory, m-estimation function and other methods are proposed, such as maximum total correntropy criterion (MTC) [8] algorithm, total least mean m-estimate (TLMM) [9] algorithm, which have a better performance under the EIV model and can effectively improve the robustness of the algorithm to impulsive noise.

Unfortunately, all the above algorithms are designed based on an ideal Gaussian noise environment. In practice, the noise is diverse and complex, and the traditional Gaussian noise assumption has limitations. To solve this problem, in recent years, the MTGC [10] algorithm on the basis of the generalized correlation entropy theory [11] has been proposed. The MTGC algorithm can not only deal with systems whose input signals contain noise, but is also robust to generalized Gaussian noise. To further improve the convergence of the MTGC algorithm, the GMBZTC algorithm has been proposed in [12]. However, the MTGC and GMBZTC algorithms require too many parameters to be set. It is well known that the adjustment of parameters is a challenging task in practical engineering, and a huge amount of parameters require to be manually adjusted, which undoubtedly increases the difficulty of using the algorithm and the cost of the work.

Moreover, the arctangent function [13][14] was demonstrated to enhance the convergence of the algorithm. Therefore, based on logistic distance metric theory [15], a new cost function is designed and the TACLDM algorithm is proposed in this paper. The algorithm reduces two parameters that need to be set manually compared to the GMBZTC algorithm, the steady-state error is also markedly decreased in the EIV model disturbed by generalized Gaussian noise. Furthermore, it is very difficult to complete the theoretical derivation for the algorithm in the EIV model with generalized Gaussian noise interference, so the papers [10] and [12] both performed very complicated integral solving when analyzing the theoretical performance of the algorithm. It is worth mentioning that the local extreme point of the algorithm is analyzed using an elegant method that avoids the complexity of the integral solution, and the steady-state MSD is calculated. In addition, we

This work was partially supported by National Natural Science Foundation of China (grant: 62171388, 61871461, 61571374).

Haiquan Zhao, Yi Peng and Zian Cao are with the School of Electrical Engineering, Southwest Jiaotong University, Chengdu, 610031, China. (e-mail: hqzhao_swjtu@126.com; pengyi1007@163.com; ziancao_swjtu@126.com).

Corresponding author: Haiquan Zhao.

reviewed a large amount of literature and found that current algorithms designed based on the TLS method generally lack the analysis of mean-square stability. Therefore, the range of step that guarantees the stability of the algorithm is computed from the perspective of mean-square stability, and this conclusion is demonstrated by computer simulation. In conclusion, computer simulations demonstrate that the TACLDM algorithm outperforms other competing algorithms and rigorous theoretical calculations. The main contributions of this paper are summarized as follows:

1) We propose the TACLDM algorithm in this paper, which achieves better performance compared to other competing algorithms with fewer parameters in the EIV model disturbed by generalized Gaussian noise.

2) To analyze the local behavior of the TACLDM algorithm more easily, an interesting method that avoids complex integral operations is used to verify the local extreme point of the TACLDM algorithm. In addition, the range of step size that guarantees the stable operation of the TACLDM algorithm is derived and the theoretical value of the steady-state MSD is calculated.

3) In this paper, the step size interval to guarantee the stability of the TACLDM algorithm is obtained from the perspective of mean square stability, which fills the gap in the existing literature on the lack of mean-square stability analysis for TLS-based adaptive filtering algorithms.

4) The TACLDM algorithm is used for system identification and acoustic echo cancellation (AEC) application. It is demonstrated through simulation that the TACLDM algorithm has superior performance compared to other competing algorithms.

The brief description of the remaining sections of this paper is as follows. Section II briefly reviews the EIV model. A detailed derivation of the TACLDM algorithm is given in Section III. The local performance as well as the mean-square stability of the TACLDM algorithm are analyzed in detail in Section IV. Section V computes the steady-state MSD of the TACLDM algorithm. The excellent performance of the TACLDM algorithm and the strict theoretical derivation of this paper are demonstrated by computer simulations in Section VI. Section VII provides a conclusion.

## II. SYSTEM MODEL

A linear model as illustrated in **Fig. 1**, can be represented as $d(\tau) = \boldsymbol{\omega}_o^T \boldsymbol{x}(\tau)$ where $\boldsymbol{\omega}_o \in \mathbb{R}^{L \times 1}$ denotes the unknown weight vector, $d(\tau) \in \mathbb{R}$ and $\boldsymbol{x}(\tau) \in \mathbb{R}^{L \times 1}$ are the desired signal and input vector.

Both input and output vectors disturbed by noise are represented in the known EIV models [6] as

$$\bar{\boldsymbol{x}}(\tau) = \boldsymbol{x}(\tau) + \boldsymbol{u}(\tau) \quad (1)$$

$$\bar{d}(\tau) = d(\tau) + v(\tau) \quad (2)$$

where $\boldsymbol{u}(\tau) \in \mathbb{R}^{L \times 1}$ and $v(\tau) \in \mathbb{R}$ are the input noise with variance $\sigma_i^2$ and the output noise with variance $\sigma_o^2$, respectively. Moreover, $\boldsymbol{u}(\tau)$ are supposed to be independently and identically distributed.

The actual output signal $y(\tau)$ and error signal $e(\tau)$ and can be indicated as

$$y(\tau) = \boldsymbol{\omega}(\tau)^T \bar{\boldsymbol{x}}(\tau) \quad (3)$$

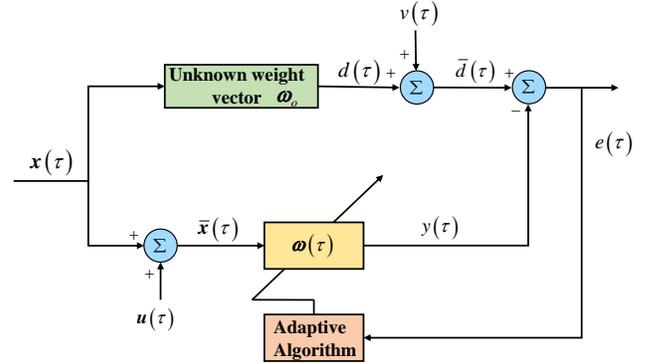

**Fig. 1.** EIV model

$$e(\tau) = \tilde{d}(\tau) - \boldsymbol{\omega}(\tau)^T \bar{\boldsymbol{x}}(\tau) \quad (4)$$

where $\boldsymbol{\omega}(\tau) \in \mathbb{R}^{L \times 1}$ is the weight vector.

## III. THE TACLDM ALGORITHM DESCRIPTION

To address the issue of too many parameters of the existing algorithms and to further improve the convergence performance by the arctangent framework. Inspired by the theory of logical distance metrics [15], the cost function of the TACLDM algorithm is proposed as

$$J_{TACLDM}(\boldsymbol{\omega}) = E\left\{\arctan\left(1\bigg/2\gamma\left[\cosh\left(\frac{e(\tau)}{\gamma\|\bar{\boldsymbol{\omega}}(\tau)\|}\right)+1\right]\right)\right\} \quad (5)$$

where $\bar{\boldsymbol{\omega}}(\tau) \stackrel{def}{=} \left[\sqrt{\varepsilon}, -\boldsymbol{\omega}(\tau)^T\right]^T$, $\varepsilon \stackrel{def}{=} \sigma_o^2/\sigma_i^2$ denotes the noise variance ratio. $\gamma$ is a positive real number which is logistic kernel and $\cosh(x) = \dfrac{\exp(x) + \exp(-x)}{2}$. Then, the gradient vector $g_{TACLDM}(\boldsymbol{\omega})$ is obtained by taking the derivative of $J_{TACLDM}(\boldsymbol{\omega})$

$$\mathbf{g}_{TACLDM}(\boldsymbol{\omega}) = \frac{\partial J_{TACLDM}(\boldsymbol{\omega})}{\partial \boldsymbol{\omega}(\tau)}$$

$$= E\left\{\frac{\sinh\left(\dfrac{e(\tau)}{\gamma\|\bar{\boldsymbol{\omega}}\|}\right)\left[\|\bar{\boldsymbol{\omega}}\|\bar{\boldsymbol{x}} + \dfrac{e(\tau)\boldsymbol{\omega}}{\|\bar{\boldsymbol{\omega}}\|}\right]}{2\gamma^2\|\bar{\boldsymbol{\omega}}\|^2\left[\cosh\left(\dfrac{e(\tau)}{\gamma\|\bar{\boldsymbol{\omega}}\|}\right)+1\right]^2\left(1+\psi(\tau)^2\right)}\right\} \quad (6)$$

$$\psi(\tau) = 1\bigg/2\gamma\left[\cosh\left(\frac{e(\tau)}{\gamma\|\bar{\boldsymbol{\omega}}(\tau)\|}\right)+1\right] \quad (7)$$

where $\sinh(x) = \dfrac{\exp(x) - \exp(-x)}{2}$.

The instantaneous gradient vector $\hat{\mathbf{g}}_{TACLDM}(\boldsymbol{\omega})$ of the

TACLDM algorithm can be obtained by (6), i.e.

$$\hat{\mathbf{g}}_{TACLDM}(\boldsymbol{\omega}) = \frac{\sinh\left(\frac{e(\tau)}{\gamma\|\bar{\boldsymbol{\omega}}\|}\right)\left[\|\bar{\boldsymbol{\omega}}\|\bar{\mathbf{x}} + \frac{e(\tau)\boldsymbol{\omega}}{\|\bar{\boldsymbol{\omega}}\|}\right]}{2\gamma^2\|\bar{\boldsymbol{\omega}}\|^2\left[\cosh\left(\frac{e(\tau)}{\gamma\|\bar{\boldsymbol{\omega}}\|}\right)+1\right]^2\left(1+\psi(\tau)^2\right)} \quad (8)$$

The weight vector update formula is given by gradient ascent method as

$$\boldsymbol{\omega}(\tau+1) = \boldsymbol{\omega}(\tau) + \mu\hat{\mathbf{g}}_{TACLDM}(\boldsymbol{\omega})$$

$$= \boldsymbol{\omega}(\tau) + \mu\frac{\sinh\left(\frac{e(\tau)}{\gamma\|\bar{\boldsymbol{\omega}}\|}\right)\left[\|\bar{\boldsymbol{\omega}}\|\bar{\mathbf{x}} + \frac{e(\tau)\boldsymbol{\omega}}{\|\bar{\boldsymbol{\omega}}\|}\right]}{2\gamma^2\|\bar{\boldsymbol{\omega}}\|^2\left[\cosh\left(\frac{e(\tau)}{\gamma\|\bar{\boldsymbol{\omega}}\|}\right)+1\right]^2\left(1+\psi(\tau)^2\right)} \quad (9)$$

*Remark1:* References [10][12], MTGC and GMBZTC algorithms have a lot of parameters to be adjusted in their weight vector update formulas, which brings great difficulties in practical use. From (9), the algorithm proposed in this paper only needs to adjust the step parameter $\mu$ and logistic kernel $\gamma$.

To describe the improvement of the algorithm by arctangent framework, we have compared the incorporation of the arctangent function and the absence of the arctangent function, which are illustrated in **Fig. 2**, **Fig. 3**, respectively. The parameters are selected as $\sigma_i^2 = \sigma_o^2 = 0.1$, $L = 2$, $\mathbf{R} = \mathbf{I}$, $\boldsymbol{\omega}_o = [-0.6, 0.8]^T$. Obviously, the upward trend of **Fig. 2** is steeper, so the arctangent framework can effectively enhance the convergence behavior of TACLDM algorithm.

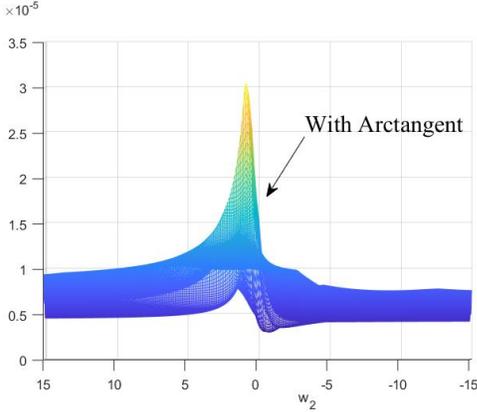

**Fig. 2.** cost function with arctangent framework

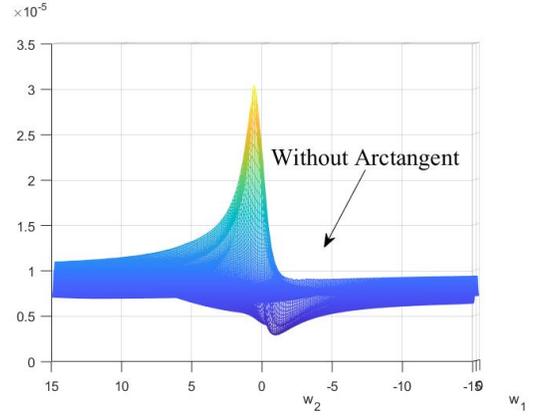

**Fig. 3.** cost function without arctangent framework

Finally, Table I illustrates the exact steps of the TACLDM algorithm.

TABLE I
Summary of TACLDM Algorithm

**Initialization:** $\boldsymbol{\omega}(0) = \mathbf{0}$

**Parameters:** $\gamma$  $\mu$

**For** $\tau = 0, 1, 2$

$$e(\tau) = \tilde{d}(\tau) - \boldsymbol{\omega}(\tau)^T \bar{\mathbf{x}}(\tau)$$

$$\bar{\boldsymbol{\omega}}(\tau) \stackrel{def}{=} \left[\sqrt{\varepsilon}, -\boldsymbol{\omega}(\tau)^T\right]^T$$

$$\psi(\tau) = 1 \bigg/ 2\gamma\left[\cosh\left(\frac{e(\tau)}{\gamma\|\bar{\boldsymbol{\omega}}(\tau)\|}\right)+1\right]$$

$$\hat{\mathbf{g}}_{TACLDM}(\boldsymbol{\omega}) = \frac{\sinh\left(\frac{e(\tau)}{\gamma\|\bar{\boldsymbol{\omega}}\|}\right)\left[\|\bar{\boldsymbol{\omega}}\|\bar{\mathbf{x}} + \frac{e(\tau)\boldsymbol{\omega}}{\|\bar{\boldsymbol{\omega}}\|}\right]}{2\gamma^2\|\bar{\boldsymbol{\omega}}\|^2\left[\cosh\left(\frac{e(\tau)}{\gamma\|\bar{\boldsymbol{\omega}}\|}\right)+1\right]^2\left(1+\psi(\tau)^2\right)}$$

$$\boldsymbol{\omega}(\tau+1) = \boldsymbol{\omega}(\tau) + \mu\hat{\mathbf{g}}_{TACLDM}(\boldsymbol{\omega})$$

**END**

## IV. PERFORMANCE ANALYSIS

### A. Local Extreme Point

In this section, to better derive the local behavior of the TACLDM algorithm, we make some similar assumptions [16][17][18].

*Assumption 1*: The matrix $\mathbf{R} = E[\mathbf{x}(\tau)\mathbf{x}^T(\tau)]$ is positive definite and full rank.

*Assumption 2*: The input noise $\mathbf{u}(\tau)$ and output noise $v(\tau)$ are both zero-mean generalized Gaussian noise. In addition, $\mathbf{u}(\tau)$, $v(\tau)$, and $\mathbf{x}(\tau)$ are separate from each other.

*Assumption 3*: At convergence to steady state, $\sinh\left(\frac{e(\tau)}{\gamma\|\bar{\boldsymbol{\omega}}\|}\right)$ and $\mathbf{x}(\tau)$ are uncorrelated.

***Assumption 4***: The variance of the noise $u(\tau)$ and $v(\tau)$ are very small. At convergence to steady state, the $\sinh(\cdot)$ and $\cosh(\cdot)$ functions are simplified using the Taylor approximation as follows

$$\sinh(\eta) \approx \eta$$
$$\cosh(\eta) \approx 1 \qquad (10)$$

The gradient vector of the algorithm at $\boldsymbol{\omega}_o$, can be expressed as

$$\mathbf{g}_{TACLDM}(\boldsymbol{\omega}_o) = E\left\{ \frac{\sinh\left(\frac{e_{\boldsymbol{\omega}_o}(\tau)}{\gamma \|\bar{\boldsymbol{\omega}}_o\|}\right)\left[\|\bar{\boldsymbol{\omega}}_o\|\bar{x} + \frac{e_{\boldsymbol{\omega}_o}(\tau)\boldsymbol{\omega}_o}{\|\bar{\boldsymbol{\omega}}_o\|}\right]}{2\gamma^2 \|\bar{\boldsymbol{\omega}}_o\|^2 \left[\cosh\left(\frac{e_{\boldsymbol{\omega}_o}(\tau)}{\gamma \|\bar{\boldsymbol{\omega}}_o\|}\right) + 1\right]^2 (1 + \psi(\tau)^2)} \right\} \quad (11)$$

where $e_{\boldsymbol{\omega}_o}(\tau) = \bar{d}(\tau) - \boldsymbol{\omega}_o^T \bar{x}(\tau) = v(\tau) - \boldsymbol{\omega}_o^T u(\tau)$ is estimation error and $\|\bar{\boldsymbol{\omega}}_o\|^2 = \|\boldsymbol{\omega}_o\|^2 + \varepsilon$.

For further derivation, we perform the following calculation, which clearly holds when $\eta$ belongs to a zero-mean generalized Gaussian distribution

$$\begin{aligned}
&COV(\sinh(\eta), \cosh(\eta)) \\
&= E\{\sinh(\eta)\cosh(\eta)\} - E\{\sinh\eta\}E\{\cosh\eta\} \\
&= E\{\sinh(2\eta)\} - 0 \\
&= 0
\end{aligned} \qquad (12)$$

From (12), (11) can be obtained as

$$\mathbf{g}_{TACLDM}(\boldsymbol{\omega}_o) = \frac{E\left\{\sinh\left(\frac{e_{\boldsymbol{\omega}_o}(\tau)}{\gamma \|\bar{\boldsymbol{\omega}}_o\|}\right)\left[\|\bar{\boldsymbol{\omega}}_o\|\bar{x} + \frac{e_{\boldsymbol{\omega}_o}(\tau)\boldsymbol{\omega}_o}{\|\bar{\boldsymbol{\omega}}_o\|}\right]\right\}}{E\left\{2\gamma^2 \|\bar{\boldsymbol{\omega}}_o\|^2 \left[\cosh\left(\frac{e_{\boldsymbol{\omega}_o}(\tau)}{\gamma \|\bar{\boldsymbol{\omega}}_o\|}\right) + 1\right]^2 (1 + \psi(\tau)^2)\right\}} \qquad (13)$$

Under **A3**, at point $\boldsymbol{\omega}_o$, the molecular part of (13) can be further obtain as

$$E\left\{\sinh\left(\frac{e_{\boldsymbol{\omega}_o}(\tau)}{\gamma \|\bar{\boldsymbol{\omega}}_o\|}\right)\left[\|\bar{\boldsymbol{\omega}}_o\|\bar{x} + \frac{e_{\boldsymbol{\omega}_o}(\tau)\boldsymbol{\omega}_o}{\|\bar{\boldsymbol{\omega}}_o\|}\right]\right\}$$
$$= E\left[\sinh\left(\frac{e_{\boldsymbol{\omega}_o}(\tau)}{\gamma \|\bar{\boldsymbol{\omega}}_o\|}\right)\right]E\left[\|\bar{\boldsymbol{\omega}}_o\|x\right] + E\left[\sinh\left(\frac{e_{\boldsymbol{\omega}_o}(\tau)}{\gamma \|\bar{\boldsymbol{\omega}}_o\|}\right) \frac{v(\tau)\boldsymbol{\omega}_o + \varepsilon u(\tau)}{\|\bar{\boldsymbol{\omega}}_o\|}\right] \qquad (14)$$

By the properties of the sinh function, the left part of (14) can be represented as

$$\sinh\left(\frac{e_{\boldsymbol{\omega}_o}(\tau)}{\gamma \|\bar{\boldsymbol{\omega}}_o\|}\right) = \sinh\left(\frac{v - \boldsymbol{\omega}_o^T u}{\gamma \|\bar{\boldsymbol{\omega}}_o\|}\right)$$
$$= \sinh\left(\frac{v}{\gamma \|\bar{\boldsymbol{\omega}}_o\|}\right)\cosh\left(\frac{\boldsymbol{\omega}_o^T u}{\gamma \|\bar{\boldsymbol{\omega}}_o\|}\right) - \cosh\left(\frac{v}{\gamma \|\bar{\boldsymbol{\omega}}_o\|}\right)\sinh\left(\frac{\boldsymbol{\omega}_o^T u}{\gamma \|\bar{\boldsymbol{\omega}}_o\|}\right) \qquad (15)$$

Under **A2**, taking the mathematical expectation on both sides of (15) gives

$$E\left[\frac{e_{\boldsymbol{\omega}_o}(\tau)}{\gamma \|\bar{\boldsymbol{\omega}}_o\|}\right] = E\left[\sinh\left(\frac{v}{\gamma \|\bar{\boldsymbol{\omega}}_o\|}\right)\right]E\left[\cosh\left(\frac{\boldsymbol{\omega}_o^T u}{\gamma \|\bar{\boldsymbol{\omega}}_o\|}\right)\right]$$
$$- E\left[\cosh\left(\frac{v}{\gamma \|\bar{\boldsymbol{\omega}}_o\|}\right)\right]E\left[\sinh\left(\frac{\boldsymbol{\omega}_o^T u}{\gamma \|\bar{\boldsymbol{\omega}}_o\|}\right)\right] \qquad (16)$$

To simplify (16), (17)(18) can be obtained under generalized Gaussian noise [11] as

$$E\left[\sinh\left(\frac{v}{\gamma \|\bar{\boldsymbol{\omega}}_o\|}\right)\right] = \frac{\alpha}{2\beta \Gamma(1/\alpha)} \int_{-\infty}^{+\infty} \sinh\left(\frac{v}{\gamma \|\bar{\boldsymbol{\omega}}_o\|}\right)$$
$$\times \exp\left[-\left(\frac{|v|}{\beta}\right)^\alpha\right] dv = 0 \qquad (17)$$

$$E\left[\sinh\left(\frac{\boldsymbol{\omega}_o^T u}{\gamma \|\bar{\boldsymbol{\omega}}_o\|}\right)\right] = 0 \qquad (18)$$

Substituting (17)(18) into (16) yields

$$E\left[\sinh\left(\frac{e_{\boldsymbol{\omega}_o}(\tau)}{\gamma \|\bar{\boldsymbol{\omega}}_o\|}\right)\right] = 0 \qquad (19)$$

For further derivation, the following calculations was performed using **A4**.

$$COV\left(\frac{e_{\boldsymbol{\omega}_o}(\tau)}{\gamma \|\bar{\boldsymbol{\omega}}_o\|}, \frac{v(\tau)\boldsymbol{\omega}_o + \varepsilon u(\tau)}{\|\bar{\boldsymbol{\omega}}_o\|}\right)$$
$$\approx E\left\{\frac{v(\tau) - \boldsymbol{\omega}_o^T u(\tau)}{\gamma \|\bar{\boldsymbol{\omega}}_o\|} \frac{v(\tau)\boldsymbol{\omega}_o + \varepsilon u(\tau)}{\|\bar{\boldsymbol{\omega}}_o\|}\right\}$$
$$- E\left\{\frac{v(\tau) - \boldsymbol{\omega}_o^T u(\tau)}{\gamma \|\bar{\boldsymbol{\omega}}_o\|}\right\} E\left\{\frac{v(\tau)\boldsymbol{\omega}_o + \varepsilon u(\tau)}{\|\bar{\boldsymbol{\omega}}_o\|}\right\} \qquad (20)$$
$$= \frac{\sigma_o^2 \boldsymbol{\omega}_o - \frac{\sigma_o^2}{\sigma_i^2}\sigma_i^2 \boldsymbol{\omega}_o}{\gamma \|\bar{\boldsymbol{\omega}}_o\|^2} - 0$$
$$= 0$$

To prove strictly that the right part of (14) is equal to 0, the $\sinh(\cdot)$ function should be expanded to infinite terms as

$$\sinh\left(\frac{e_{\boldsymbol{\omega}_o}(\tau)}{\gamma \|\bar{\boldsymbol{\omega}}_o\|}\right) = \sum_{n=0}^{\infty} \frac{\left(\frac{e_{\boldsymbol{\omega}_o}(\tau)}{\gamma \|\bar{\boldsymbol{\omega}}_o\|}\right)^{2n+1}}{(2n+1)!}, \frac{e_{\boldsymbol{\omega}_o}(\tau)}{\gamma \|\bar{\boldsymbol{\omega}}_o\|} \in (-\infty, +\infty) \qquad (21)$$

Then, from (20)(21) the right part of (14) can be represented as

$$E\left[\sinh\left(\frac{e_{\boldsymbol{\omega}_o}(\tau)}{\gamma \|\bar{\boldsymbol{\omega}}_o\|}\right)\frac{v(\tau)\boldsymbol{\omega}_o + \varepsilon u(\tau)}{\|\bar{\boldsymbol{\omega}}_o\|}\right]$$
$$= E\left(\sum_{n=0}^{\infty} \frac{\left(\frac{v(\tau) - \boldsymbol{\omega}_o^T u(\tau)}{\gamma \|\bar{\boldsymbol{\omega}}_o\|}\right)^{2n+1}}{(2n+1)!}\right) E\left(\frac{v(\tau)\boldsymbol{\omega}_o + \varepsilon u(\tau)}{\|\bar{\boldsymbol{\omega}}_o\|}\right) \qquad (22)$$
$$= 0$$

Substituting (19)(22) into (14), $\mathbf{g}_{TACLDM}(\boldsymbol{\omega}_o) = 0$ can be easily gotten. Then, to prove that $\boldsymbol{\omega}_o$ is an extreme point, the

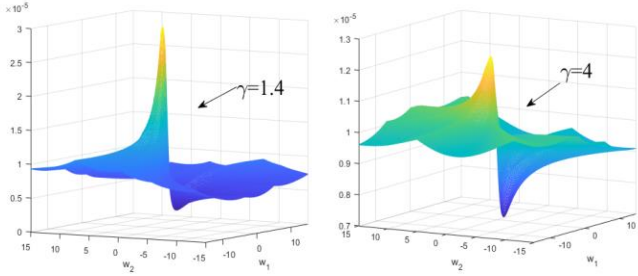

Hessian matrix $H_{TACLDM}(\boldsymbol{\omega}_o)$ needs to be computed. Firstly, the $H_{TACLDM}(\boldsymbol{\omega}_o)$ has been calculated at the bottom of page 5.

Furthermore, using **A4**, the Hessian matrix at $\boldsymbol{\omega}_o$ can be derived as

**Fig. 4** Surfaces of $J_{TACLDM}(\boldsymbol{\omega})$ with different $\gamma$

$$H_{TACLDM}(\boldsymbol{\omega}_o) = \left(-\frac{1}{8\gamma^3 \|\bar{\boldsymbol{\omega}}_o\|^2 \left(1+\frac{1}{16\gamma^2}\right)} - \frac{E\left(e_{\boldsymbol{\omega}_o}^2(\tau)\right)}{128\gamma^7 \|\bar{\boldsymbol{\omega}}_o\|^4 \left(1+\frac{1}{16\gamma^2}\right)^2}\right) B(\tau) \quad (23)$$

$$B(\tau) = E\left(\bar{\boldsymbol{x}}(\tau)\bar{\boldsymbol{x}}^T(\tau) + \frac{2e_{\boldsymbol{\omega}_o}(\tau)\bar{\boldsymbol{x}}\boldsymbol{\omega}_o^T}{\|\bar{\boldsymbol{\omega}}_o\|^2} + \frac{e_{\boldsymbol{\omega}_o}^2(\tau)\boldsymbol{\omega}_o\boldsymbol{\omega}_o^T}{\|\bar{\boldsymbol{\omega}}_o\|^4}\right) \quad (24)$$

Based on **A1** and **A2**, the right side of (24) can be computed as

$$E[\bar{\boldsymbol{x}}(\tau)\bar{\boldsymbol{x}}^T(\tau)] = E[\boldsymbol{x}(\tau)\boldsymbol{x}^T(\tau)] + \sigma_i^2 = \boldsymbol{R} + \sigma_i^2 \boldsymbol{I} \quad (25)$$

$$E[e_{\boldsymbol{\omega}_o}(\tau)\bar{\boldsymbol{x}}(\tau)] = -E[\boldsymbol{\omega}_o^T \boldsymbol{u}(\tau)\boldsymbol{u}(\tau)] = -\sigma_i^2 \boldsymbol{\omega}_o \quad (26)$$

$$E[e_{\boldsymbol{\omega}_o}^2(\tau)] = E[v^2 + \boldsymbol{u}^T(\tau)\boldsymbol{\omega}_o^T \boldsymbol{\omega}_o \boldsymbol{u}(\tau)] = \sigma_i^2 \|\bar{\boldsymbol{\omega}}_o\|^2 \quad (27)$$

where $\boldsymbol{I}$ is the identity matrix.

Substituting (25)(26)(27) into (23) yields

$$H_{TACLDM}(\boldsymbol{\omega}_o) = \frac{-16\gamma^4 - \gamma^2 - \sigma_i^2}{128\gamma^7 \|\bar{\boldsymbol{\omega}}_o\|^2 \left(1+\frac{1}{16\gamma^2}\right)^2} \left[\boldsymbol{R} + \left(\sigma_i^2 \boldsymbol{I} - \sigma_i^2 \frac{\boldsymbol{\omega}_o \boldsymbol{\omega}_o^T}{\|\bar{\boldsymbol{\omega}}_o\|^2}\right)\right] \quad (28)$$

Obviously, from formula (28), it can be seen that the matrix $H_{TACLDM}(\boldsymbol{\omega}_o)$ is negative definite, thus $\boldsymbol{\omega}_o$ is a local maximum point.

The surfaces of $J_{TACLDM}(\boldsymbol{\omega})$ for different $\gamma$ are shown in **Fig. 4**. From **Fig. 4**, there is only one locally maximal point $\boldsymbol{\omega} = \boldsymbol{\omega}_o$ in the wide range of $\boldsymbol{\omega}$, which ensures that there is only a unique solution in the neighborhood of $\boldsymbol{\omega}_o$ in the wide range of $\boldsymbol{\omega}$.

### B. Mean Convergence

This subsection derives the step size intervals that guarantee the stable operation of the TACLDM algorithm from the perspective of mean convergence.

The gradient error arises due to the fact that the instantaneous value replaces the expected value in the updating formula (9) for the weights. Thus, the gradient error is expressed as

$$z(\boldsymbol{\omega}) = \hat{\boldsymbol{g}}_{TACLDM}(\boldsymbol{\omega}) - \boldsymbol{g}_{TACLDM}(\boldsymbol{\omega}) \quad (29)$$

Introducing formula (29) into (9), we have

$$\boldsymbol{\omega}(\tau+1) = \boldsymbol{\omega}(\tau) + \mu \boldsymbol{g}_{TACLDM}(\boldsymbol{\omega}) + \mu z(\boldsymbol{\omega}) \quad (30)$$

Define $\Delta\boldsymbol{\omega}(\tau) = \boldsymbol{\omega}_o - \boldsymbol{\omega}(\tau)$, then (30) can be denoted as

$$\Delta\boldsymbol{\omega}(\tau+1) = \Delta\boldsymbol{\omega}(\tau) - \mu[\boldsymbol{g}_{TACLDM}(\boldsymbol{\omega}) + z(\boldsymbol{\omega})] \quad (31)$$

Through $\boldsymbol{g}_{TACLDM}(\boldsymbol{\omega}) \approx -H_{TACLDM}(\boldsymbol{\omega}_o)\Delta\boldsymbol{\omega}(\tau)$ [12], (31) can be approximated as

$$\Delta\boldsymbol{\omega}(\tau+1) \approx [\mu H_{TACLDM}(\boldsymbol{\omega}_o) + \boldsymbol{I}]\Delta\boldsymbol{\omega}(\tau) - \mu z(\boldsymbol{\omega}) \quad (32)$$

Taking the mathematical expectation on both sides of (32), yields

$$E[\Delta\boldsymbol{\omega}(\tau+1)] \approx [\mu H_{TACLDM}(\boldsymbol{\omega}_o) + \boldsymbol{I}] E[\Delta\boldsymbol{\omega}(\tau)] \quad (33)$$

Based on (33), it's easy to see that to ensure local convergence of the TACLDM algorithm is that all eigenvalues of the matrix $\boldsymbol{I} + \mu H_{TACLDM}(\boldsymbol{\omega}_o)$ should satisfy

$$|\boldsymbol{I} + \mu \lambda_{\max}\{H_{TACLDM}(\boldsymbol{\omega}_o)\}| < 1 \quad (34)$$

where $\lambda_{\max}\{H_{TACLDM}(\boldsymbol{\omega}_o)\}$ is maximum eigenvalue of the matrix $H_{TACLDM}(\boldsymbol{\omega}_o)$. Finally, the range of $\mu$ that allow the TACLDM algorithm to operate stably can be deduced as

$$0 < \mu < \frac{256\gamma^7 \|\bar{\boldsymbol{\omega}}_o\|^2 \left(1+\frac{1}{16\gamma^2}\right)^2}{\left(-16\gamma^4 - \gamma^2 - \sigma_i^2\right)\lambda_{\max}\left\{\boldsymbol{R} + \left(\sigma_i^2 \boldsymbol{I} - \sigma_i^2 \frac{\boldsymbol{\omega}_o \boldsymbol{\omega}_o^T}{\|\bar{\boldsymbol{\omega}}_o\|^2}\right)\right\}} \quad (35)$$

### C. Mean Square Convergence Analysis

In this subsection, we try to analyze the TACLDM algorithm from the perspective of mean-square stability and derive the corresponding step size range.

Substituting $\Delta\boldsymbol{\omega}(\tau) = \boldsymbol{\omega}_o - \boldsymbol{\omega}(\tau)$ into (9) yields

$$\Delta\boldsymbol{\omega}(\tau+1) = \Delta\boldsymbol{\omega}(\tau) - \mu\hat{\boldsymbol{g}}_{TACLDM}(\boldsymbol{\omega}) \quad (36)$$

Then, taking the Euclidean-norm and expectation on both sides of (36)

---

$$H_{TACLDM}(\boldsymbol{\omega}) = \frac{\partial \boldsymbol{g}_{TACLDM}}{\partial \boldsymbol{\omega}^T}$$

$$= E\left\{\frac{2+\cosh\left(\frac{e_{\boldsymbol{\omega}_o}(\tau)}{\gamma\|\bar{\boldsymbol{\omega}}_o\|}\right) - \cosh^2\left(\frac{e_{\boldsymbol{\omega}_o}(\tau)}{\gamma\|\bar{\boldsymbol{\omega}}_o\|}\right)}{2\gamma^3 \|\bar{\boldsymbol{\omega}}_o\|^4 \left[\cosh\left(\frac{e_{\boldsymbol{\omega}_o}(\tau)}{\gamma\|\bar{\boldsymbol{\omega}}_o\|}\right)+1\right]^3 \left(1+\frac{1}{16\gamma^2}\right)} + \frac{\sinh^2\left(\frac{e_{\boldsymbol{\omega}_o}(\tau)}{\gamma\|\bar{\boldsymbol{\omega}}_o\|}\right)}{4\gamma^5 \|\bar{\boldsymbol{\omega}}_o\|^4 \left[\cosh\left(\frac{e_{\boldsymbol{\omega}_o}(\tau)}{\gamma\|\bar{\boldsymbol{\omega}}_o\|}\right)+1\right]^5 \left(1+\frac{1}{16\gamma^2}\right)^2}\right\} E\left\{\left[\|\bar{\boldsymbol{\omega}}_o\|\bar{\boldsymbol{x}} + \frac{e(\tau)\boldsymbol{\omega}}{\|\bar{\boldsymbol{\omega}}_o\|}\right]\left[-\bar{\boldsymbol{x}}^T \|\bar{\boldsymbol{\omega}}_o\| - \frac{e(\tau)\boldsymbol{\omega}^T}{\|\bar{\boldsymbol{\omega}}_o\|}\right]\right\}$$

$$MSD(\tau+1) = MSD(\tau)$$
$$+ \mu^2 E\left[\hat{\mathbf{g}}_{TACLDM}^T(\boldsymbol{\omega})\hat{\mathbf{g}}_{TACLDM}(\boldsymbol{\omega})\right] \quad (38)$$
$$- 2\mu E\left[\Delta\boldsymbol{\omega}^T(\tau)\hat{\mathbf{g}}_{TACLDM}(\boldsymbol{\omega})\right]$$

where $MSD(\tau) \stackrel{def}{=} E\|\Delta\boldsymbol{\omega}(\tau)\|^2$, the convergence of TACLDM algorithm satisfies

$$MSD(\tau) - MSD(\tau+1) > 0 \quad (39)$$

Substituting (39) into (38) yields

$$0 < \mu < \frac{2E\left[\Delta\boldsymbol{\omega}^T(\tau)\hat{\mathbf{g}}_{TACLDM}(\boldsymbol{\omega})\right]}{E\left[\hat{\mathbf{g}}_{TACLDM}^T(\boldsymbol{\omega})\hat{\mathbf{g}}_{TACLDM}(\boldsymbol{\omega})\right]} \quad (40)$$

At convergence to steady state, based on **A3** and **A4**, expectations of the molecule and denominator in (40) can be calculated as [19]

$$2E\left[\Delta\boldsymbol{\omega}^T(\tau)\hat{\mathbf{g}}_{TACLDM}(\boldsymbol{\omega}_o)\right] = \frac{\sigma_i^2}{4\gamma^3\left(1+\frac{1}{16\gamma^2}\right)^2} \quad (41)$$

$$E\left[\hat{\mathbf{g}}_{TACLDM}^T(\boldsymbol{\omega}_o)\hat{\mathbf{g}}_{TACLDM}(\boldsymbol{\omega}_o)\right]$$
$$= \frac{E\left[e_{\boldsymbol{\omega}_o}^2 \bar{\mathbf{x}}^T \bar{\mathbf{x}}\|\bar{\boldsymbol{\omega}}_o\|^4 + 2e_{\boldsymbol{\omega}_o}(k)\bar{\mathbf{x}}\boldsymbol{\omega}_o^T\|\bar{\boldsymbol{\omega}}_o\|^2 + e_{\boldsymbol{\omega}_o}^4 \boldsymbol{\omega}_o^T\boldsymbol{\omega}_o\right]}{64\gamma^6\|\bar{\boldsymbol{\omega}}_o\|^8\left(1+\frac{1}{16\gamma^2}\right)^2} \quad (42)$$

The molecule in (42) is further calculated as
$$E\left[e_{\boldsymbol{\omega}_o}^2 \bar{\mathbf{x}}^T \bar{\mathbf{x}}\|\bar{\boldsymbol{\omega}}_o\|^4\right] = \mathbf{x}^T(\tau)\mathbf{x}(\tau)\sigma_i^2\|\bar{\boldsymbol{\omega}}_o\|^6 + \sigma_i^2\sigma_o^2\|\bar{\boldsymbol{\omega}}_o\|^4$$
$$+ \boldsymbol{\omega}_o^T\boldsymbol{\omega}_o(L+2)\sigma_i^4\|\bar{\boldsymbol{\omega}}_o\|^4 \quad (43)$$

$$2E\left[e_{\boldsymbol{\omega}_o}(\tau)\bar{\mathbf{x}}\boldsymbol{\omega}_o^T\|\bar{\boldsymbol{\omega}}_o\|^2\right] = -6\boldsymbol{\omega}_o^T\boldsymbol{\omega}_o\sigma_i^4\|\bar{\boldsymbol{\omega}}_o\|^4 \quad (44)$$

$$E\left[e_{\boldsymbol{\omega}_o}^4 \boldsymbol{\omega}_o^T\boldsymbol{\omega}_o\right] = 3\sigma_i^4\|\bar{\boldsymbol{\omega}}_o\|^4 \boldsymbol{\omega}_o^T\boldsymbol{\omega}_o \quad (45)$$

Substituting (41)(42)(43)(44)(45) into (40) gives

$$0 < \mu < \frac{16\gamma^3\|\bar{\boldsymbol{\omega}}_o\|^4}{L\sigma_x^2\|\bar{\boldsymbol{\omega}}_o\|^2 + \sigma_o^2 + \boldsymbol{\omega}_o^T\boldsymbol{\omega}_o(L-1)\sigma_i^2} \quad (46)$$

After combining (35) and (46), the step size interval that ensures the stability of the TACLDM algorithm is denoted as

$$0 < \mu < \min\left(\frac{256\gamma^7\|\bar{\boldsymbol{\omega}}_o\|^2\left(1+\frac{1}{16\gamma^2}\right)^2}{\left(-16\gamma^4 - \gamma^2 - \sigma_i^2\right)\lambda_{\max}\left\{R + \left(\sigma_i^2 I - \sigma_i^2 \frac{\boldsymbol{\omega}_o\boldsymbol{\omega}_o^T}{\|\bar{\mathbf{w}}_o\|^2}\right)\right\}},\right.$$
$$\left.\frac{16\gamma^3\|\bar{\boldsymbol{\omega}}_o\|^4}{L\sigma_x^2\|\bar{\boldsymbol{\omega}}_o\|^2 + \sigma_o^2 + \boldsymbol{\omega}_o^T\boldsymbol{\omega}_o(L-1)\sigma_i^2}\right) \quad (47)$$

## V. STEADY-STATE MSD

In this subsection, the steady-state MSD of the TACLDM algorithm is computed under the condition of generalized Gaussian noise. At the steady state $\boldsymbol{\omega} \approx \boldsymbol{\omega}_o$, then (29) can be approximated as $z(\boldsymbol{\omega}) \approx \hat{\mathbf{g}}_{TACLDM}(\boldsymbol{\omega})$. Further, squaring both sides of (32) and taking the Euclidean norm as

$$E\left[\|\Delta\boldsymbol{\omega}(\tau+1)\|_Q^2\right] \approx E\left[\|\Delta\boldsymbol{\omega}(\tau)\|_T^2\right] + \mu^2 E\left[\|z(\boldsymbol{\omega})\|_Q^2\right] \quad (48)$$

where non-negative definite matrix $Q$ is a freely selectable, and $T$ is expressed as

$$T \stackrel{def}{=} \left[I + \mu H_{TACLDM}(\boldsymbol{\omega}_o)\right] Q \left[I + \mu H_{TACLDM}(\boldsymbol{\omega}_o)\right] \quad (49)$$

Using vectorize operation $vec\{\bullet\}$, which is to stack the columns of its matrix argument into a single column vector, we define $q = vec\{Q\}$, $m = vec\{M\}$, and $t = vec\{T\}$.

First, we give the formula for computing the matrix $M$ as

$$M \stackrel{def}{=} E\left[z(\boldsymbol{\omega}_o)z^T(\boldsymbol{\omega}_o)\right]$$
$$= \frac{\sigma_i^2}{64\gamma^6\|\bar{\boldsymbol{\omega}}_o\|^2\left(1+\frac{1}{16\gamma^2}\right)^2}\left[R + \sigma_i^2 I - \sigma_i^2\frac{\boldsymbol{\omega}_o\boldsymbol{\omega}_o^T}{\|\bar{\boldsymbol{\omega}}_o\|^2}\right] \quad (50)$$

Furthermore, using the relationship between the matrix trace [20] and the vectorization operator $vec\{\bullet\}$, as well as $Q$ is deterministic and symmetric, we can get

$$E\left[\|z(\boldsymbol{\omega}_o)\|_Q^2\right] = E\left[tr\{Q(z(\boldsymbol{\omega}_o)z^T(\boldsymbol{\omega}_o))\}\right] = tr\{QM\} = m^T q \quad (51)$$

Substituting (51) into (48), we have
$$E\left[\|\Delta\boldsymbol{\omega}(\tau+1)\|_q^2\right] \approx E\left[\|\Delta\boldsymbol{\omega}(\tau)\|_t^2\right] + \mu^2 m^T q \quad (52)$$

Due to the relationship between the vectorize operation and the Kronecker product and (49), $t = vec\{T\}$ can be expressed as
$$t = Fq \quad (53)$$

where
$$F = \left(I + \mu H_{TACLDM}(\boldsymbol{\omega}_o)\right) \otimes \left(I + \mu H_{TACLDM}(\boldsymbol{\omega}_o)\right) \quad (54)$$

At steady state, substituting (54) into (52) gives
$$E\left[\|\Delta\boldsymbol{\omega}(\infty)\|_q^2\right] \approx E\left[\|\Delta\boldsymbol{\omega}(\infty)\|_{Fq}^2\right] + \mu^2 m^T q \quad (55)$$

From (55), we have
$$E\left[\|\Delta\boldsymbol{\omega}(\infty)\|_{(I-F)q}^2\right] \approx \mu^2 m^T q \quad (56)$$

Now, $q$ is selected as
$$q = (I - F)^{-1} vec\{I\} \quad (57)$$

Then, the steady-state MSD can be gotten as
$$E\left[\|\Delta\boldsymbol{\omega}(\infty)\|^2\right] \approx \mu^2 m^T (I - F)^{-1} vec\{I\} \quad (58)$$

## VI. SIMULATION

In this section, to demonstrate the performance of the TACLDM algorithm, all simulated values are averaged by 1000 separate Monte Carlo (MC) runs. The performance has been evaluated for different non-Gaussian/impulsive background noise. Unless otherwise indicated, the filter order is set to 9, the number of samples is 3000, and the input and output noise were zero-mean Gaussian with $\sigma_i^2 = \sigma_o^2 = 0.1$.

The normalized MSD (NMSD) is used to measure the performance of all algorithms and is expressed as

$$NMSD = 10\log\left[\left\|\boldsymbol{\omega}(\tau) - \boldsymbol{\omega}_o\right\|^2 / \left\|\boldsymbol{\omega}_o\right\|^2\right] \quad (59)$$

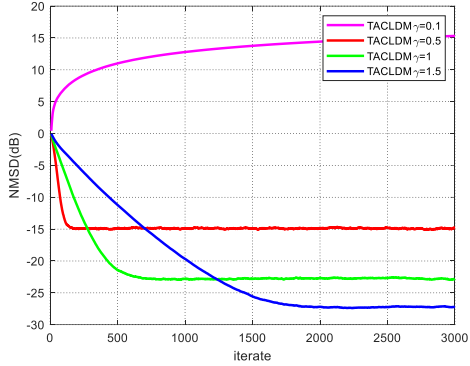

**Fig. 5.** TACLDM algorithm curve under different $\gamma$

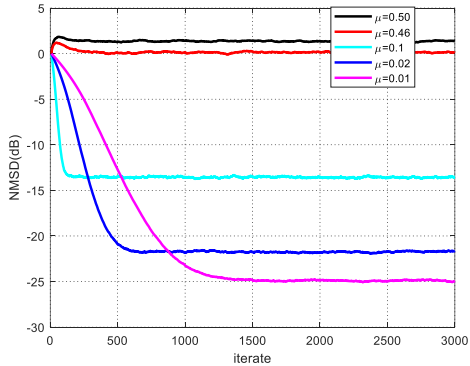

**Fig. 6.** TACLDM algorithm curve under different $\mu$

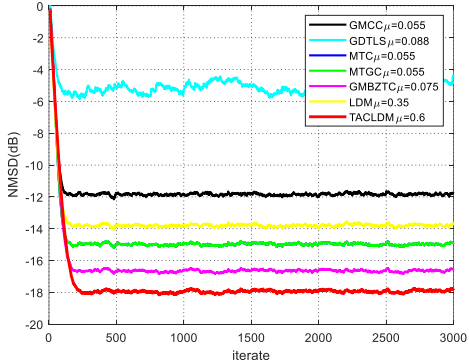

**Fig. 7.** Under Gaussian noise and impulsive noise

It is worth mentioning that to ensure fairness in comparisons between different algorithms, the parameter values of the different algorithms should be chosen in a way that guarantees that all algorithms show the equal initial convergence.

*A. The effect of $\gamma$ on the TACLDM algorithm*

In this section, to obtain the effect of $\gamma$ on the TACLDM algorithm, the step size is $\mu = 0.1$. The impact of various $\gamma$ on the TACLDM algorithm is illustrated in **Fig. 5**.

As shown in **Fig. 5**, with increasing $\gamma$, the algorithm has a smaller NMSD, but slower convergence. A smaller $\gamma$ yields faster convergence, but when the $\gamma$ is too small so that the $\mu$ is no longer in the range of stable operation, then the algorithm diverges.

*B. The effect of $\mu$ on the TACLDM algorithm*

In this section, to obtain the effect of $\mu$ on the TACLDM algorithm, the step size is $\gamma = 0.5$. The impact of various $\mu$ on the TACLDM algorithm is shown in **Fig. 6**.

As illustrated in **Fig. 6**, with increasing $\mu$, the convergence speed of the algorithm increases, but the NMSD also rises, while the $\mu$ decreases, the convergence speed of the algorithm degrades and the NMSD degrades as well. Moreover, when the $\mu$ reaches the critical value of 0.46, the algorithm is in a critical state. When the step size $\mu$ exceeds 0.46 and reaches 0.5, the TACLDM algorithm has diverged.

*C. Performance Comparison*

**1) Under Gaussian and impulsive noise environment**

As indicated in **Fig. 7**, the $u(\tau)$ is zero mean Gaussian noise, and the $v(\tau)$ a mixture of zero-mean Gaussian noise and impulsive noise with the probability of 0.01. Here, $\gamma_{TACLDM} = \gamma_{LDM} = 1.4$, $\gamma = 1$, $\alpha = 2$, and $\sigma = 1$. Moreover, an appropriate $\mu$ is chosen so that all algorithms have the same initial convergence rate.

As shown in **Fig. 7**, the TACLDM algorithm has better performance compared to other competing algorithms. When the $v(\tau)$ contains impulsive noise interference, there is a significant performance degradation of GDTLS, while the TACLDM algorithm still performs well. This is because the $\mathbf{g}_{TACLDM}$ of the TACLDM algorithm approximated to be 0 when $e(\tau)$ is very large, which eliminates the case of incorrectly updating the weights for various outliers, including impulsive noise. In addition, MTGC shrinks to MTC at $\alpha = 2$.

**2) Under Laplacian noise environment**

As indicated in **Fig. 8**, the $u(\tau)$ is zero mean Gaussian noise with variance is $\sigma_i^2 = 0.1$, and the $v(\tau)$ is zero mean Laplacian noise with variance is $\sigma_o^2 = 1$. Choose $\gamma_{TACLDM} = \gamma_{LDM} = 1.3$, $\alpha = 1.56$, $\gamma = 0.1$, $\sigma = 1$ and suitable $\mu$ for different algorithms so that different algorithms have the same initial convergence rate. **Fig. 8** shows the algorithm proposed in this paper obviously outperforms other algorithms in the Laplace environment.

**3) Under Uniform and impulsive noise environment**

As depicted in **Fig. 9**, the $u(\tau)$ is the uniform noise in the range $\left[-\sqrt{2}, \sqrt{2}\right]$, and the output noise $v(\tau)$ is uniform noise in the range $\left[-\sqrt{2}, \sqrt{2}\right]$ mixture of impulsive noise with the probability of 0.01. Choose $\gamma_{TACLDM} = \gamma_{LDM} = 4$  $\alpha = 6$,

$\gamma = 0.5$, $\sigma = 1$. From **Fig. 9**, the TACLDM algorithm still performs very well under uniform noise and obviously has lower steady state error compared to other competing algorithms.

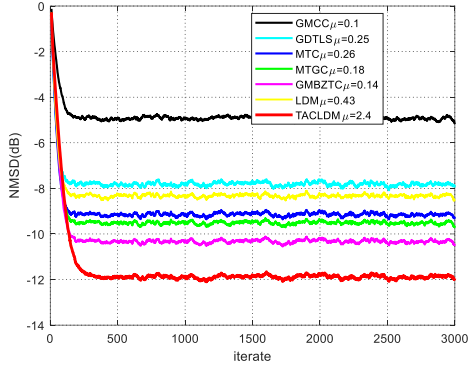

**Fig. 8.** Under Laplacian noise

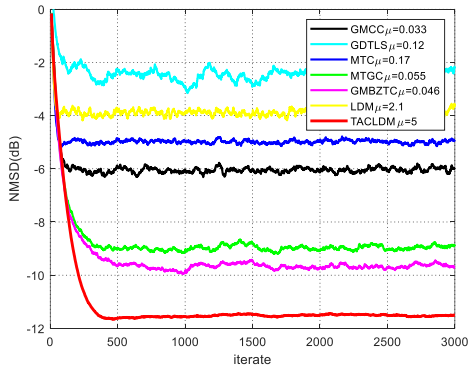

**Fig. 9.** Under Uniform noise

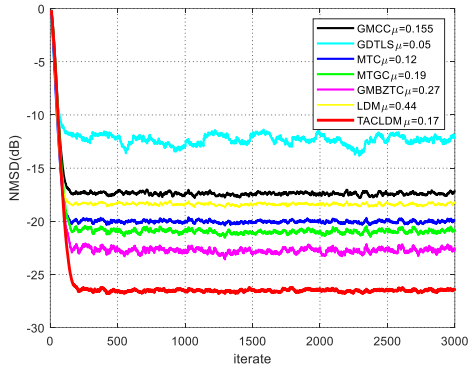

**Fig. 10.** Under binary noise

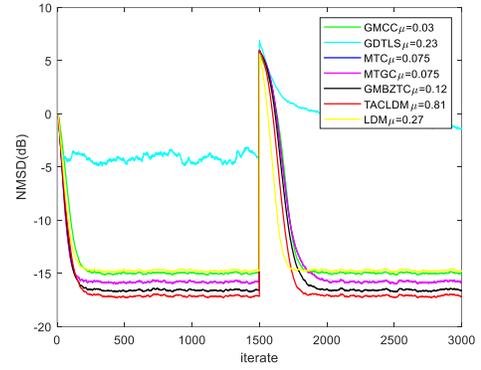

**Fig. 11** Tracking performance under Gaussian noise

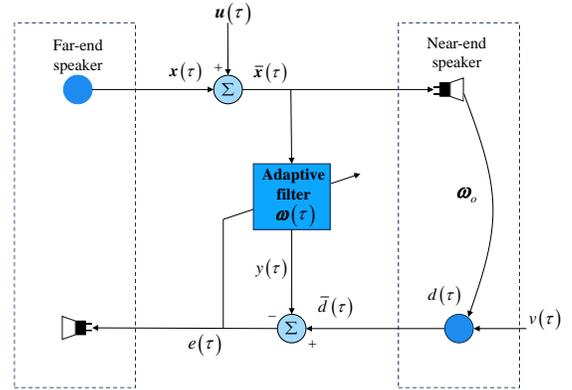

**Fig. 12.** AFA use in AEC

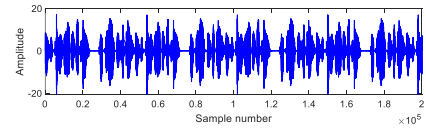

**Fig. 13** Speech input vector

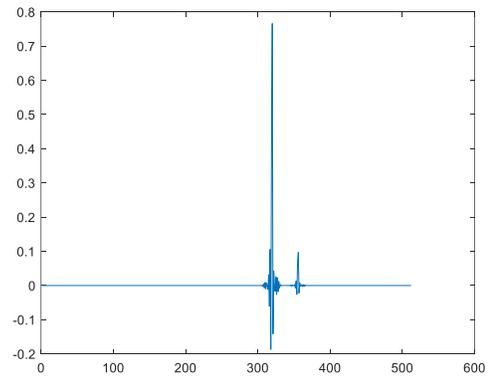

**Fig. 14** Echo channel

**4) Under binary noise**

As depicted in **Fig. 10**, both $u(\tau)$ and $v(\tau)$ are binary noise in the range of $[-0.1, 0.1]$ and $\mathrm{P}\{x = -1\} = 0.5$, $\mathrm{P}\{x = 1\} = 0.5$. Choose $\gamma_{TACLDM} = \gamma_{LDM} = 0.72$, $\alpha = 2.34$, $\gamma = 0.91$, $\sigma = 0.48$. From **Fig. 10,** the TACLDM algorithm still obviously outperforms other competing algorithms under binary noise.

**5) The tracking performance**

In **Fig. 11**, at the 1500-th input samples, $\omega_o$ becomes to $-\omega_o$. Both $u(\tau)$ and $v(\tau)$ are zero mean Gaussian with variance is $\sigma_i^2 = \sigma_o^2 = 0.1$, and $v(\tau)$ mixes impulsive noise with the probability of 0.01. Here, $\gamma_{TACLDM} = \gamma_{LDM} = 1.2$, $\alpha = 2$, $\gamma = 1$, and $\sigma = 1$. As indicated in **Fig. 11**, the tracking behavior of the TACLDM algorithm performs markedly better than other competing algorithms.

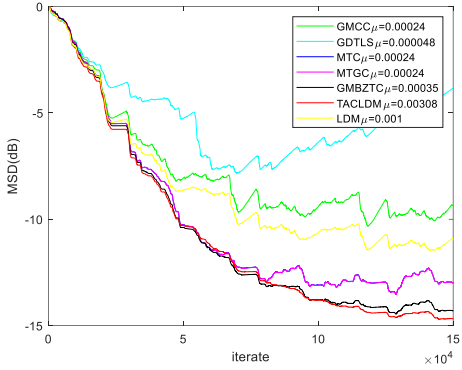

**Fig. 15.** Comparison of different algorithms under AEC applications

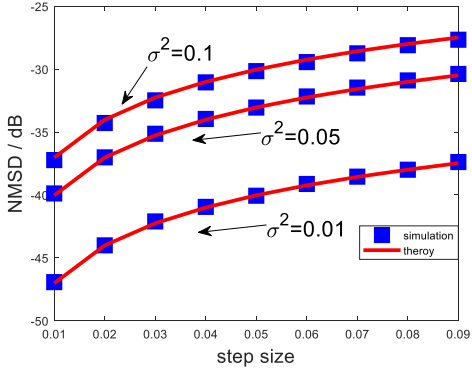

**Fig. 16.** Under Gaussian noise

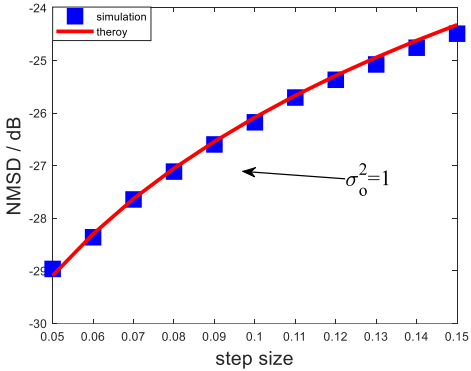

**Fig. 17.** Under Laplacian noise

**6) Application in AEC**

**Fig. 12** illustrates the AEC model, **Fig. 13** illustrates the real speech input vector, while **Fig. 14** presents the echo signal. The $x(\tau)$ from the far-end speaker is an input vector and is interfered with by $u(\tau)$, and the output vector $\bar{x}(\tau)$ is the near-end speaker input. $d(\tau)$ is generated by the echo channel $\omega_o$. After the adaptive filtering algorithm, $\omega(\tau)$ will eventually fit with $w_o$. Finally, the $d(\tau)$ after the interference of noise $v(\tau)$ is subtracted from the echo signal $y(\tau)$ and transmitted to the far-end speaker to achieve the purpose of echo cancellation.

In **Fig. 15**, TACLDM algorithm is compared with other algorithms under AEC application. Both $u(\tau)$ and $v(\tau)$ are zero mean Gaussian with variance is $\sigma_i^2 = \sigma_o^2 = 0.1$, and $v(\tau)$ mixes impulsive noise with the probability of 0.01. Select $\gamma_{TACLDM} = \gamma_{LDM} = 1.1$, $\alpha = 2$, $\gamma = 1$ and $\sigma = 1$. As depicted in **Fig. 15**, there is a significant performance degradation of the GDTLS algorithm due to the output noise containing impulsive noise. Whereas the LDM algorithm is unable to handle the interference of input noise, the GMCC is less effective in handling the input noise. In addition, the curve of MTGC overlaps with MTC due to $\alpha = 2$. Moreover, the TACLDM algorithm has better performance than the other algorithms and uses fewer parameters compared to GMBZTC.

*B. Verification of theoretical steady state MSD*

In this subsection, our calculation of the theoretical steady-state MSD of the TACLDM algorithm will be verified by computer simulation.

In **Fig. 16**, both $u(\tau)$ and $v(\tau)$ are zero mean Gaussian with variance is $\sigma^2 = \sigma_i^2 = \sigma_o^2 = 0.1$ and $\gamma_{TACLDM} = 1.4$. As shown in **Fig. 16**, the theoretical steady-state MSD of the TACLDM algorithm and the simulated values are in good agreement for different conditions.

To better verify the theoretical calculations under generalized Gaussian noise, the steady-state MSD comparison under Laplace noise is also derived in **Fig. 17**. In **Fig. 17**, the input noise $u(\tau)$ is zero mean Gaussian noise with $\sigma_i^2 = 0.1$, and the output noise $v(\tau)$ is zero mean Laplacian noise with variance is $\sigma_o^2 = 1$. Choose $\gamma_{TACLDM} = 1.3$. From **Fig.17**, the simulated values also agree very well with the theoretical values under Laplace noise, further proving the correctness of the derivation of the TACLDM algorithm under generalized Gaussian noise.

## VII. CONCLUSION

In this paper, a new cost function is designed based on the LDM theory and arctangent framework, and the TACLDM algorithm is proposed. The new algorithm obtains superior performance using fewer parameters compared to other competing algorithms under the EIV model when both input and output signals are disturbed by generalized Gaussian noise. In addition, an interesting method is used to avoid complex integral solutions, and the proposed algorithm is also analyzed from the perspective of mean-square stability in this paper, which fills the gap of the lack of mean-square stability analysis in the existing literature on TLS-based adaptive filtering algorithms. Finally, the correctness of the algorithm derivation and the superiority of the algorithm performance are verified by computer simulation.